\documentclass{elsart}
\usepackage{natbib}
\usepackage{epsfig}
\usepackage{unites2e}
\usepackage{amsmath}
\usepackage{amssymb}
\def\etall{{\it et al.}}

\begin{document}
\runauthor{Colin}
\begin{frontmatter}
\title{Measurement of air and nitrogen fluorescence light yields induced by
electron beam for UHECR experiments }

\author[LAPP]{P. Colin}\author[UTAH]{}\footnote{contact address: {\tt colin@physics.utah.edu}}, 
\author[JINR]{A. Chukanov}, \author[JINR]{V. Grebenyuk},
\author[JINR]{D. Naumov}\author[INFN]{},
\author[LAPP]{P. N\'ed\'elec}, \author[JINR]{Y. Nefedov},
\author[LIP] {A. Onofre}, \author[JINR]{S. Porokhovoi}, \author[JINR]{B. Sabirov,}
\author[JINR]{L. Tkatchev}

\textbf{The MACFLY Collaboration:}

\address[LAPP]{Laboratoire d'Annecy-le-Vieux de Physique des Particules, IN2P3/CNRS, \\  Universit\'e de Savoie, Annecy-le-Vieux, FRANCE}
\address[JINR]{Joint Institute for Nuclear Research \\ Dubna, Moscow region, RUSSIA}
\address[LIP] {Laboratorio de Instrumenta\c{c}\~{a}o e Fisica Experimental de Particulas\\ Coimbra, PORTUGAL}
\address[UTAH]{Department of Physics, University of Utah\\ Salt-Lake-City, Utah, USA}
\address[INFN]{Instituto Nazionale di Fisica Nucleare, Firenze, ITALY}

\begin{abstract}
Most of the Ultra High Energy Cosmic Ray (UHECR) experiments and
projects (HiRes, AUGER, TA, EUSO, TUS,...) use air fluorescence to
detect and measure extensive air showers (EAS). The precise
knowledge of the Fluorescence Light Yield (FLY) is of paramount
importance for the reconstruction of UHECR. The MACFLY -
Measurement of Air Cherenkov and Fluorescence Light Yield -
experiment has been designed to perform such FLY measurements. In
this paper we will present the results of FLY in the 290-440~nm
wavelength range for dry air and pure nitrogen, both excited by
electrons with energy of 1.5~MeV, 20~GeV and 50~GeV. The
experiment uses a $^{90}Sr$ radioactive source for low energy
measurement and a CERN SPS e$^-$ beam for high energy. We find
that the FLY is proportional to the deposited energy ($E_d$) in
the gas and we show that the air fluorescence properties remain
constant independently of the electron energy. At the reference
point: atmospheric dry air at 1013~hPa and 23$^\circ$C, the ratio
$FLY/E_d$=17.6~photon/MeV with a systematic error of 13.2\%.
\end{abstract}
\end{frontmatter}

\section{Introduction}
The physics of the Ultra High Energy Cosmic Rays (UHECR) is a
challenging search in the field of cosmic ray physics. There is a
special interest to measure the energy spectrum in the region of
$10^{20}$~eV where the GZK cutoff is predicted~\cite{GZK}. For more
than 40 years, many experiments looked for such UHECR events but
only a few dozen of them have been recorded so far. The lack of
statistics and the discrepancies between the results of the two main
experiments AGASA~\cite{agasa} and HIRES~\cite{hires}, requires new
experiments to solve the UHECR and the GZK puzzle. In the coming
years, with the deployment of ground based detectors: the Pierre
Auger Observatory~\cite{Auger} and the Telescope Array~\cite{TA},
this number of events will increase dramatically, reaching several
hundreds. In the longer term, space-based experiments (OWL, EUSO,
TUS)~\cite{euso} are foreseen, with the goal of reaching a thousand
events per year.

For most of the past, present and future experiments, the
detection technique relies, at least, on the measurement of the
air fluorescence produced by the incoming cosmic ray showering in
the atmosphere. This light, observed in the near-UV region, is
induced by the de-excitation of the molecule of air (mainly
$N_{2}$) occurring along the development of the extensive air
showers (EAS). This Fluorescence Light Yield (FLY) is rather weak
($\sim$ 4 photons/m emitted in 4$\pi$), and depends significantly
on the atmospheric conditions (temperature, pressure, gas
components: H$_{2}$O, Ar, ...).

Since the 1960's,  when it was first proposed to use air
fluorescence for UHECR's detection, the FLY has been studied. In
1967, A. N. Bunner summarized the existing data in his
thesis~\cite{Bunner}. At that time the uncertainty on the FLY was
$\sim$30 \% ! Later, the discrepancy between the AGASA and HiRes
experiments led the UHECR community to pursue their efforts on the
FLY measurement. In 1996, in the context of HiRes and Telescope
Array, a new set of results~\cite{Kakimoto} was published. More
recently, M. Nagano~\etall~ released new measurements on the FLY
with systematic uncertainty of $\sim$13.2 \%~\cite{Nagano}.

A better understanding of the FLY, with a goal of setting the
systematic uncertainties below 10 \%, is now needed for the UHECR
experiments. Up to now only the pressure dependence was really
measured. To improve our knowledge of air FLY and its behavior with
respect to pressure, temperature, humidity, electron energy, shower
age, etc, new experiments are needed. An overview of the current
experiments can be found elsewhere~\cite{IWFM}.

This paper is organized as follows: Section 2 describes the
fluorescence mechanisms and presents a predictive model. In Section 3
the experimental setup is described. In Section 4 the
data taking and signal processing are discussed and in Section 5 the
calibration and systematic studies are presented. Finally our
results are described in Section 6 and a comparison with other
experiments are discussed in Section 7.

\section{Air fluorescence mechanisms and production model}
\label{sec-model}

The Fluorescence Light Yield depends on two competitive processes:
excitation and de-excitation of the air molecules.

In an EAS, the air is excited by the high energy charged particles
of the shower. When such a particle traverses the air, it ionizes
air and produces secondary low energy electrons which will excite
$N_{2}$ molecules in the low energy states (10-20~eV), which
fluoresce. The energy (and hence the wavelength) of the
fluorescence photons corresponds to the energy difference between
two excited states of an air molecule: the spectrum varies in a
large band, from UV to IR. In the UHECR experiments, the photons
are detected in the near UV band, which ranges in the 290-440 nm
window in MACFLY.

A molecular excited state can be defined by its molecular orbital
and by its vibrational state. The air fluorescence spectrum at
atmospheric pressure is a band spectrum dominated by 1N system of
$N_{2}^{+}$ and 2P system of $N_{2}$. 1N system corresponds to all
the transitions between $N_2^+$ orbitals $B^2\Sigma^+_u$ and
$X^2\Sigma^+_g$, and 2P system to all the transitions between
$N_{2}$ orbitals $C^3\Pi_u$ and $B^3\Pi_g$~\cite{D&O}.

During the de-excitation, there is a contest between the radiative
and the non-radiative processes, which are both characterized by a
time scale. The first one may produce photons in the experimental UV
band while the second process dissipates the energy via thermal
processes.

The mean lifetime $\tau_e$ of an excited state in non-isolated
conditions verifies:
\begin{equation}
    \frac{1}{\tau_e}=\frac{1}{\tau_{e_0}}+\sum{\frac{1}{\tau^i_{e_C}}}~,
\end{equation}
where the first term $\tau_{e_0}$ is the mean lifetime of the
isolated excited state. For instance the typical values of the 2P
system are $\sim$40 ns~\cite{Pancheshnyi}.

The non-radiative de-excitation of the molecules (quenching) comes
from collisions of the excited molecules with other air molecules.
The collision times $\tau^i_{e_C}$ of the excited state $e$ with the
molecule of type $i$ ($i$ = $N_2, O_2, H_2O,...$) are given by the
kinematic theory of gases:
\begin{equation}
    \nonumber 
    \tau^i_{e_C} = \frac{1}{P^{i}}\frac{\sqrt{\pi k T
    m_{N_2}}}{4\sigma^i_e}\sqrt{\frac{2m_i}{m_{N_2}+m_i}}
    \label{eq-tau_e_C}~,
\end{equation}
where $T$ is the gas temperature, $P^i$ is the partial pressure of
molecules $i$, $k$ is the Boltzmann constant, $m_{N_2}$ and $m_i$
are the molecular mass of $N_2$ and of molecule $i$, and
$\sigma^i_e$ is the cross section of the collision. The
$\tau^i_{e_C}$ have been measured since a long time and their
values vary inversely with the pressure $P$. At atmospheric
pressure the quenching time of the 2P system is a few ns.

One often expresses the relationship between the lifetime and the
pressure via the formula:
\begin{equation}
    \label{eq-tvsP}
    \frac{1}{\tau_e} = \frac{1}{\tau_{e_0}} (1 +
    \frac{P}{P'})~,
\end{equation}
P' being the pressure for which the quenching processes have a
collision time equal to the radiative de-excitations lifetime.


In our model, the quenching processes are characterized by the
quenching factors $k^i_e$, independent of the pressure, defined by:
$k^i_e=\frac{4\sigma^i_e}{\sqrt{\pi k T m_{N_2}}}
\sqrt{\frac{m_{N_2}+m_i}{2m_i}}$.

Assuming that $\sigma^i_e$ do not vary with temperature, implies that
the $k^i_e$ vary solely as the inverse square root of the
temperature: $k^i_{e}(T)=k^i_{e(T_0)}\sqrt{\frac{T_0}{T}}$. The
values of $k^i_{e(T_0)}$ are given for a reference temperature
$T_0$.

Thus, $\tau^i_{e_C}$ can be re-written as:
\begin{equation}\label{eq-tvsk}
    \tau^i_{e_C}(P,T) = \frac{1}{P \times f^i \times k^i_{e}(T)}~.
\end{equation}

For any excited states, $e$, of a gas at temperature $T$ and pressure
$P$ containing a fraction $f^i$ of molecules $i$, the probability
$\mathcal{Q}$ to have a radiative de-excitation is:

\begin{equation}
\nonumber 
\mathcal{Q} = \frac{1}{\tau_{e_0}}~/\frac{1}{\tau_e}
            = \frac{\tau_e}{\tau_{e_0}}
            = \frac{1}{1+P\tau_{e_0} \sum{f^i k^i_{e(T_0)}}\sqrt{\frac{T_0}{T}}}~.
\label{eq-quenching}
\end{equation}

For all the air fluorescence lines, at wavelength $\lambda$,
corresponding to an excited state $e$, one can express the
${{FLY}\over{E_d}}\big|_{\lambda}$ ratio of an arbitrary admixture
of $N_2$, $O_2$ and $H_2O$, as a function of pressure $P$ and
temperature $T$, via the formula: \noindent
\begin{equation}
{{FLY}\over{E_d}}\Big|_{\lambda} (P,T)  =  \frac{ \chi_{N_2} \cdot {
{{FLY}\over{E_d}}\Big|^{N_{2}}_{{\lambda,P=0}} }}
    {1 + P\cdot\tau_{e_{0}}
    \sum{f^i k^i_{e(T_0)}}
    \sqrt{\frac{T_0}{T}}}~,
\label{eq-melange}
\end{equation}
where ${{FLY}\over{E_d}}|^{N_2}_{\lambda,P=0}$ is the limit of the
FLY ratio for pure nitrogen at $P$=0~hPa (where no quenching
effect by molecular collisions is expected) and $\chi_{N_{2}}$ is
the mass fraction of nitrogen in the gas admixture
($i=O_2,N_2,H_2O$).

In our model we assume that the fraction of the deposited energy
in the gas converted in nitrogen excitation is $\chi_{N_{2}}$. The
parameters entering equation~\ref{eq-melange}:
${{FLY}\over{E_d}}|^{N_2}_{\lambda,P=0}$, $\tau_{e_{0}}$,
$k^{N_{2}}_{e(T_0)}$, $k^{O_{2}}_{e(T_0)}$ and
$k^{H_{2}O}_{e(T_0)}$ are given for 24 wavelengths ($\lambda$)
corresponding to 5 excited states ($e$) at the reference
temperature $T_0 = 20^\circ$C (293.15~K). The values of these
parameters come from both our MACFLY measurements and from already
published results:~\cite{Bunner}~\cite{Nagano}~\cite{D&O} for the
spectra ($\lambda$) and~\cite{Pancheshnyi}~\cite{Brunet} for the
lifetimes ($\tau_{e_0}$) and the quenching factors
($k_{e(T_0)}^i$).

In this paper, we will show experimental results for pure nitrogen
($N_2$) and for our experimental dry air (DA) --- an admixture
80\%($N_2$)-20\%($O_2$) --- as a function of the pressure. From this it
will be then possible to derive the behavior of the atmospheric dry
air (ADA) --- an admixture 78.08\%($N_2$)-0.93\%($Ar$)-20.99\%($O_2$)
--- for which we assume the FLY for argon to be identical to the FLY
of nitrogen due to its catalysis effect on nitrogen
fluorescence~\cite{grun}. Then the ADA becomes in our model
--- an admixture 79\%($N_2$)-21\%($O_2$).

Using this, we formulate the general FLY of any humid "atmospheric
air" (AA), {\it i.e.} any admixture composed of $(1-\mu)$ dry air (ADA)
and $\mu$ water vapor, where $\mu$ is the molecular fraction of
water in the admixture ($\mu=f^{H_2O}$ as defined in equation
\ref{eq-tvsk}). We get:
\begin{equation}\label{eq-anyairfly}
    \frac{FLY}{E_d}\Big|^{AA}_{\lambda}(P,T,\mu) =
    \frac{FLY}{E_d}\Big|^{ADA}_{\lambda,P=0} \times
    \frac{1-\frac{18}{29}\mu}
         {1 + P \left( \frac{1-\mu}{P'^{ADA}_{e(T_0)}}+
                       \frac{\mu}{P'^{H_{2}O}_{e(T_0)}} \right)
                       \sqrt{\frac{T_0}{T}}}~,
\end{equation}
where
\begin{equation}\label{eq-PDA'}
P'^{ADA}_{e}=\frac{1}{\tau_{e_{0}}\cdot ( 0.79 k^{N_2}_{e} + 0.21
k^{O_2}_{e})} ~~~~~\text{and}~~~~~
P'^{H_2O}_{e}=\frac{1}{\tau_{e_{0}}\cdot k^{H_2O}_{e}}~.
\end{equation}

We assume that the energy fraction transferred to ADA molecules is
proportional to the mass fraction of ADA in the admixture rather
than the molecular fraction (1$-\mu$). The numerator
(1$-\frac{18}{29} \mu$) represents this mass fraction. It is a
simplification of the exact formula $\frac{(1-\mu)\cdot
29}{(1-\mu)\cdot 29 + \mu\cdot 18}$ for $\mu\ll$1 as it is the
case in the atmospheric air.

\section{Experimental setup}
The MACFLY\footnote{Measurement of Air Cherenkov and Fluorescence
Light Yield} experiment~\cite{Colin} is twofold, it has been
designed to measure both the FLY induced by single electron track
and the FLY produced by a high energy electromagnetic shower
developing in the air. It is composed of two devices: MF1 which is
used for single track FLY measurements and MF2 which measures the
fluorescence produced by an electromagnetic shower. In this paper we
will present only the results obtained with MF1 using both a
$\beta^-$ radioactive source (Strontium 90) and an electron test
beam (CERN/SPS-X5 line). The results of the shower FLY with MF2 will
be presented elsewhere~\cite{mf2-paper}.

\begin{figure}[htb]
  \includegraphics[width=\columnwidth]{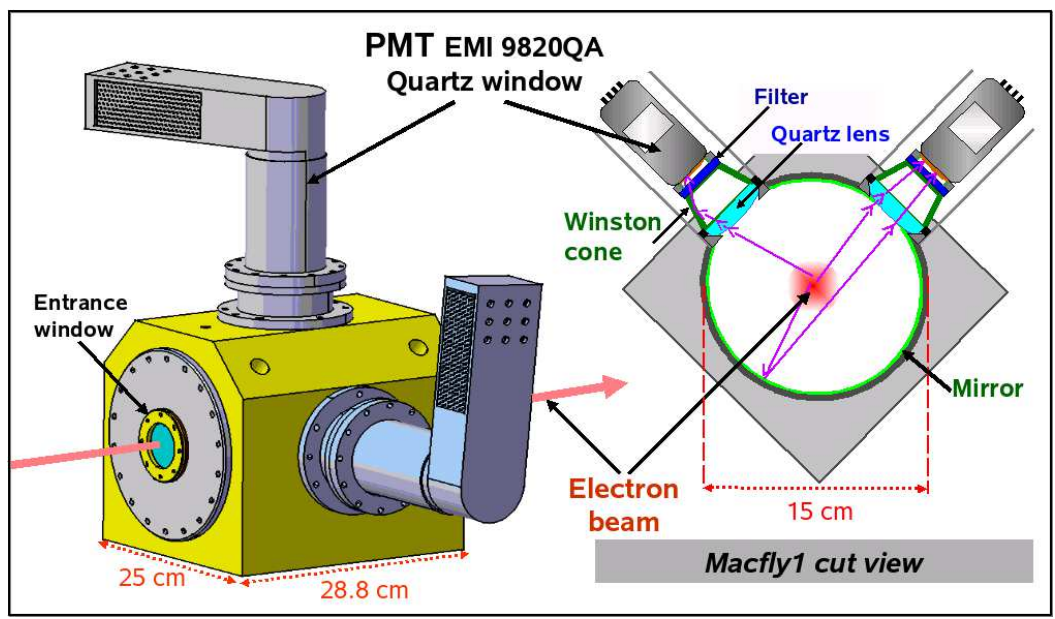}
  \caption{Schematic view of the MF1 chamber (left) and a cut view
   of its optical system (right) showing the multilayered mirror,
   the quartz lens, the Winston cones, the BG3 filters and the PMTs.}
  \label{fig-MF1_overview}
\end{figure}

The MF1 experiment is composed of a pressurized chamber (with
pressure from 0 to 1200~hPa) containing the gas, equipped with a
trigger system (see Figure~\ref{fig-MF1_overview}).
This device has an internal cylindrical chamber, 150 mm in diameter
and 288 mm long. An ionizing particle (E~$\gtrsim$ 1~MeV) reaches the
gas volume through the entrance window (0.25~mm of black Delrin),
crosses the gas volume and leaves the chamber at the exit window
(0.8~mm of aluminium) where it reaches the trigger plane.

\begin{figure}[htb]
\begin{center}
  \includegraphics[width=10cm]{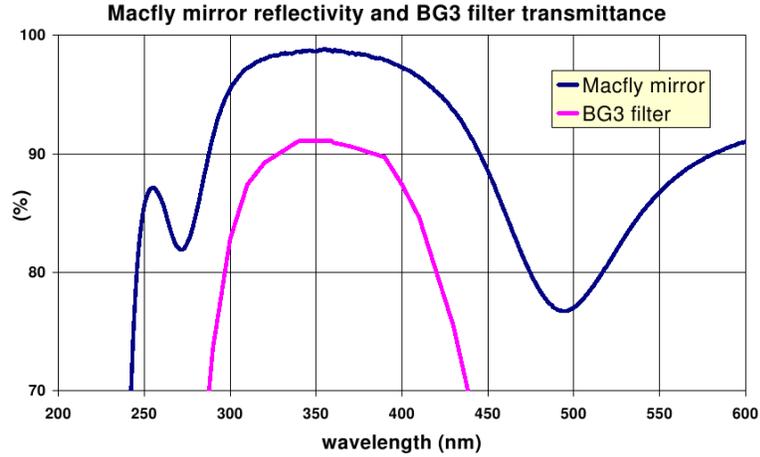}
  \caption{MF1 multilayered mirror reflectivity coefficient as a
  function of wavelength (upper curve) and BG3 UV filter
  transmission coefficient (lower curve).}
  \label{fig-Mirror}
\end{center}
\end{figure}

\begin{figure}
  \includegraphics[width=\columnwidth]{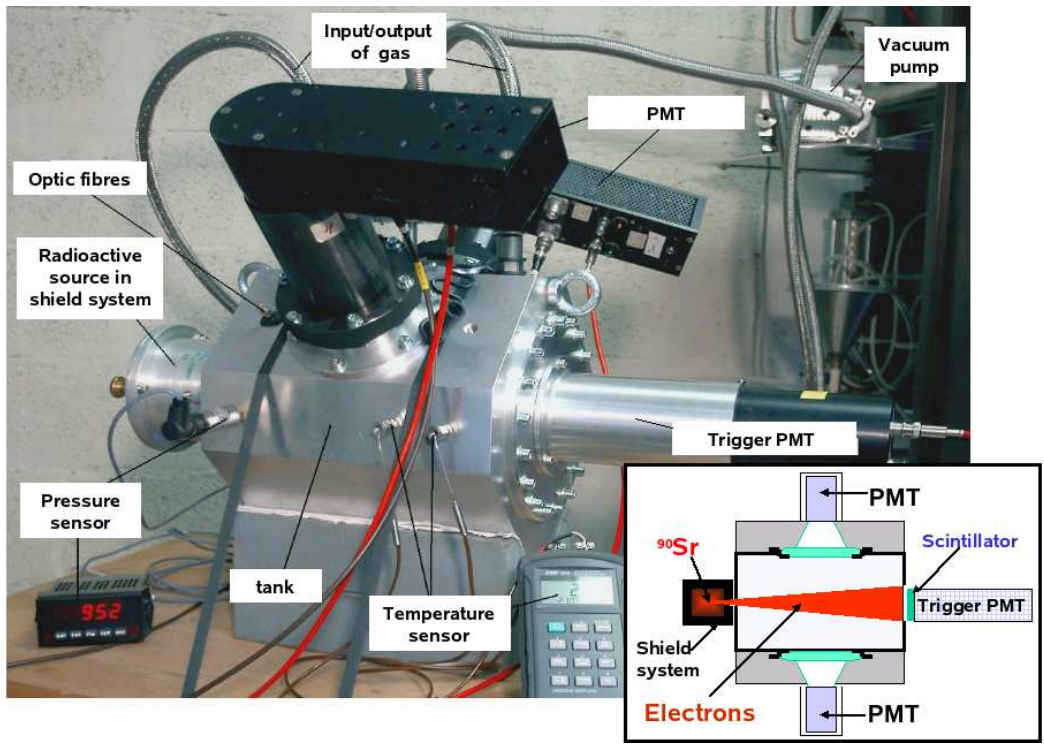}
  \caption{MF1 setup in its laboratory configuration. The radioactive
  source is embedded in a shielded box, and the trigger is mounted on
  the device itself.}
  \label{fig-MF1_LAPP}
\end{figure}

An optical system collects the induced fluorescence light, and
focuses it on two UV sensitive phototubes (PMT) EMI9820QA. Each
PMT receives photons via an optical system which is composed of a
multilayered mirror covering the chamber, a quartz lens focusing
the light, a Winston cone and wide band filters which allow the
selection of the appropriate wavelength. The multilayered
mirror\footnote{From internal to external layer: 90~nm (Al) +
43~nm ($SiO_2$) + 43~nm ($HfO_2$) + 43~nm ($SiO_2$) + 43~nm
($HfO_2$)} was designed to have its best reflectance in the
appropriate air fluorescence band (see Figure~\ref{fig-Mirror}).
The filters are Schott BG3 filters with a large band of
transmittance: 290-440~nm.
We have measured the FLY for two gases: pure nitrogen and dry air
($N_{2}$-80\%, $O_{2}$-20\%). The MF1 experiment was tested in two
different setup configurations corresponding to low and high energy.
For the low energy runs (labeled MF1-lab) a $\beta^-$ source
($^{90}Sr$) was used and the trigger system was integrated in the
overall system (see Figure~\ref{fig-MF1_LAPP}). For the high energy
runs (labeled MF1-beam) the CERN SPS XP5 beam facility was used.
In this later case the source and the trigger system are removed
from the main tank, and external systems are installed along the
beam line: a position sensitive ($\sigma \sim 0.5 ~mm$) X-Y delay
chamber upstream from the MF1 device to record both the horizontal
and the vertical beam position and two double plane trigger systems
before and after the MF1 setup. The electron beam was a pulsed beam
of about 10~000 electrons per spill (4.8~s duration) every 16.8~s
with a beam spot of about $4\times7$~mm$^2$. The electron energy was
selected in the range between 10 and 100 GeV.

\section{Data taking and signal fitting}
The data was collected on an event per event basis. For every
event the PMT signals (MF1 and trigger counters) were recorded by
a QADC (CAEN-V792) which integrated the charge during a gate of
100~ns. The beam position was also recorded for the MF1-beam
configuration. During a run, which consists of typically 10$^6$
triggers, we recorded two kinds of events: the beam events (BE),
used for the FLY measurement, and the random events (RE), when no
beam is present, for the background estimations and studies. In
MF1-lab configuration BE are triggered when an electron reaches
the trigger plane after the exit window. In MF1-beam
configuration, the electron should reach two trigger planes before
and after the MF1 chamber. In the analysis, we also suppressed BE
not detected in the central region of the delay chamber.

The FLY is rather weak ($\sim$4 ph/m) and most of the produced
photons are lost in the chamber. Among those reaching the PMT
photocathode, some are converted in photoelectron (pe). Overall, the
mean number of detected photoelectron is about 0.01 pe/event.
To extract the mean Detected Light (DL), we performed a fit of the
data by a function describing the expected signal; for every
event, a PMT detects an integer number of photon: 0, 1, 2 or more
converting them into photoelectrons. So the overall PMT
spectrum is described by a weighted sum of individual
photoelectron contributions, the weights following a Poisson
distribution.

The zero photoelectron contribution, the pedestal, is measured
experimentally.\\
The single photoelectron contribution is described by two
functions: one describing the standard photoelectron
multiplication at the first dynode, the other corresponds to the
photoelectron being inelastically back-scattered on the first
dynode~\cite{Philips}. The first function is described by a
Weibull\footnote{$Weibull(x)=c/b\cdot(x/b)^{c-1}\cdot
exp(-(x/b)^{c})$} distribution~\cite{Colin} while the second uses
an exponential law.
\\
The Multi-photoelectron contributions (n$\geq$2) are described by
gaussian distributions. The mean value $\mu_{n}$ and the variance
$\sigma^{2}_{n}$ of each distribution are determined by the single
photoelectron fit function parameters: $\mu_{n}=n \mu_{1} +
\mu_{0}$ and $\sigma^{2}_{n}=n \sigma^{2}_{1}$ where $\mu_{1}$ and
$\sigma^{2}_{1}$ are the single photoelectron distribution
(Weibull) mean value and
variance\footnote{$\mu_{1}=b\cdot\Gamma\left(
1+\frac{1}{c}\right)$ and $\sigma_{1}=b\sqrt{\Gamma\left(
1+\frac{2}{c}\right)-\left(\Gamma\left(
1+\frac{1}{c}\right)\right)^{2}}$} and $\mu_{0}$ the pedestal mean
value.



\begin{figure}[htb]
\begin{center}
  \includegraphics[width=10cm]{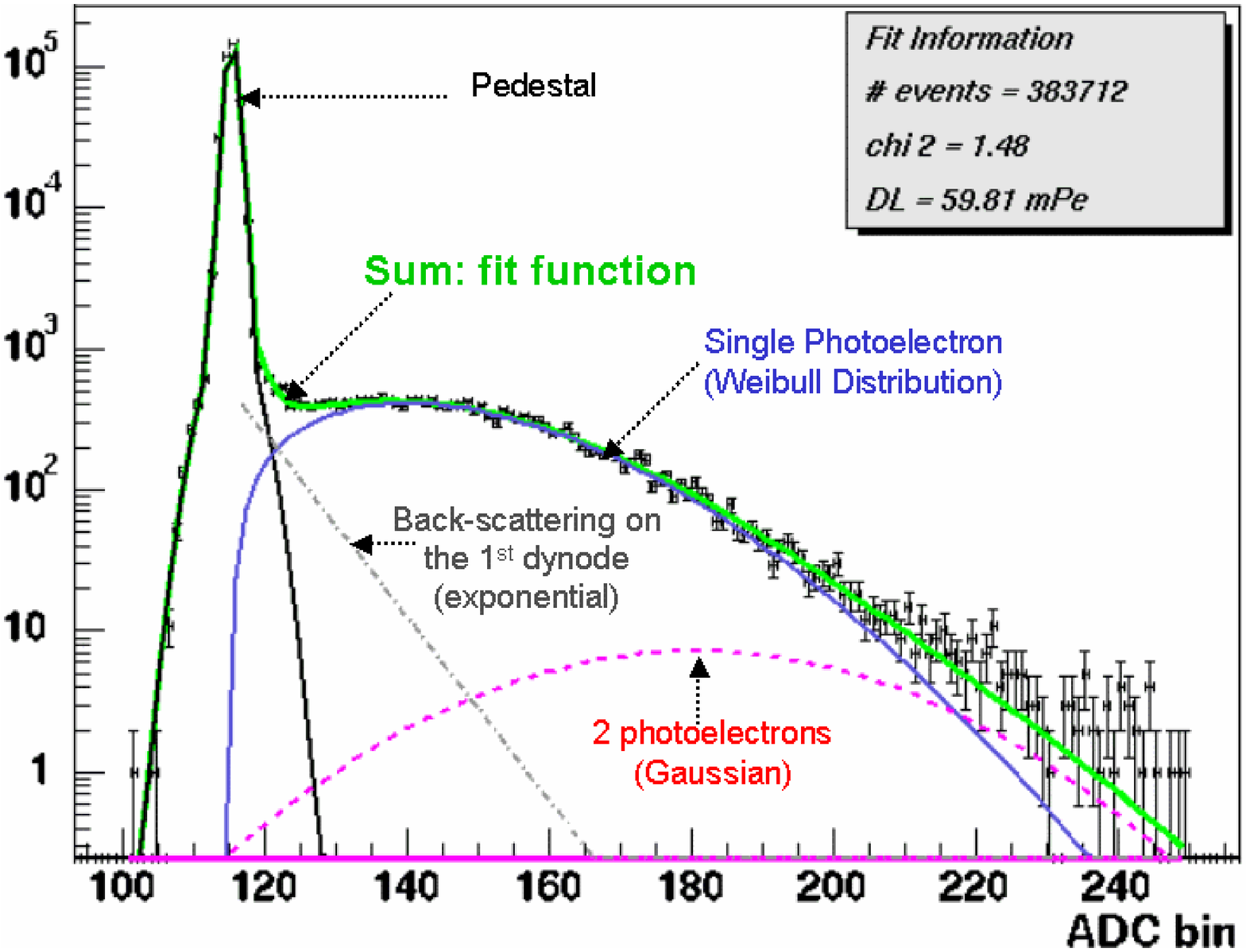}
  \caption{PMT spectrum fit function (solid line) is a sum of several
  contributions: pedestal (black solid line), single photoelectron
  (solid line), multi-photoelectron (dotted line) and back-scattered
  single  photoelectron (dotted dash line).}
  \label{fig-fit_description}
\end{center}
\end{figure}

One can see in Figure~\ref{fig-fit_description} the PMT raw signal
as well as the different contributions entering the fit function. It
is worth noting the agreement between the global fitted function and
the data extends on more than five orders of magnitude.

In the context of MACFLY data ($\sim5\times10^5$ BE/run), the low level
of PMT dark noise enables the detection of a signal as low as: 0.0002~pe/event,
which correspond to a mean sensitivity of the MF1 apparatus of about 0.04~ph/event
for the florescence light. Howerver other sources of PMT photoelectrons signal
degrade this minimal sensitivity to $\sim$0.2~ph/event.

The detected light has several origins: fluorescence light (FDL),
Cherenkov light (CDL) and background (Bgd).
\begin{equation}
    DL=FDL+CDL+Bgd~.
\end{equation}
The Bgd contribution is estimated from both the data from the RE
triggers and from BE with vacuum in the chamber (no signal
expected). \\
The Cherenkov contribution is estimated, based on a Geant4 Monte
Carlo simulation program, describing in details both the apparatus
and the interaction processes~\cite{geant4}. For MF1-lab
measurements, electrons do not produce any Cherenkov radiation. At
high energy (E $>$ 10 GeV), the Cherenkov light yield is important
but its contribution to the raw PMT signal remains small; the
Cherenkov light is emitted in the forward direction where a black
"light catcher" is installed, suppressing dramatically the reflected
Cherenkov component.

\begin{figure}[htb]
  \includegraphics[width=\columnwidth]{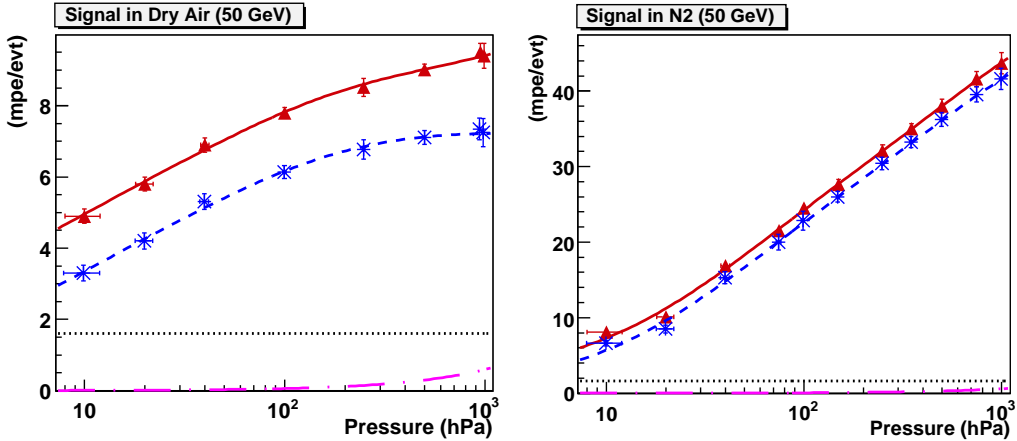}
  \caption{Measured light in dry air (left) and in pure nitrogen
(right) in milli-photoelectron per event (mpe/evt) as a function of the gas
pressure. Triangles are raw measured light data (DL);
dotted line is the Bgd estimation from vacuum measurements;
dotted dash line is CDL simulation; stars are the FDL data
(after substraction of Bgd and CDL); dashed line is a simple
FLY model fit and the solid line is its projection on DL.}
\label{fig-DL compo}
\end{figure}

Figure~\ref{fig-DL compo} shows the DL measured and the
estimation of CDL and Bgd for test beam measurements at CERN as a
function of pressure in Dry Air and nitrogen. The FDL is
determined by substraction. We can see that the main part of the
DL comes from the fluorescence.
The Fluorescence Light Yield in MF1, is determined from the
measurements from both overall and background signals and from the
estimated contribution for the Cherenkov yield:
\begin{equation}\label{eq-FLY}
FLY=\frac{DL - CDL - Bgd }{\varepsilon_{MF1}}~,
\end{equation}
where the MF1 efficiency $\varepsilon_{MF1}$ is the product of
geometrical acceptance of the detector and of the PMT quantum
efficiency ($\varepsilon_{MF1-beam}=0.00556~ph/pe$).

\section{Calibration and systematics}

The geometrical acceptance is estimated by a Geant4~\cite{geant4}
simulation program which describes in detail the apparatus. It is
able to perform the tracking of the optical photons from the
production source along the electron track to the PMT photocathodes
(Filters transmittance and Mirror reflectivity are included). The
simulation result shows that 2.75\% of the fluorescence light
emitted isotropically in the chamber reaches one PMT photocathode.

This geometrical acceptance varies with the setup configuration
(mirror, surfaces, filters,...). The optical properties of the inner
surfaces are well known: the mirror reflectance, as function of
wavelength,  was measured with less than 2\% of error. Figure
\ref{fig-Mirror} shows the result of this measurement. The other
inner surfaces were studied in laboratory with a calibration device
at 370~nm.

To check our Geant4 based simulation program, we compared the
FLY measurements for different experimental (optical system)
configurations with the simulation expectations. For instance,
having black covered some inner surfaces, the geometrical acceptance
was reduced by 24\% which was really close to the predicted
simulation value of 26\%.

Table~\ref{table-GeoAcc} summarizes the systematic relative
errors on geometrical acceptance for both experimental
configurations (MF1-lab and MF1-beam). The main contributions come
from the electron track position uncertainty (delay chamber/MF1
alignment and multiple-scattering in the chamber), from the
internal surfaces reflective properties and from the mirror
inhomogeneity. The overall geometrical acceptance uncertainty is
estimated to 7.5\% for MF1-lab and 8.2\% for MF1-beam.

\begin{table}
\begin{center}
\begin{tabular}{|l|c|c|}
  \hline
  Uncertainty sources & MF1-lab & MF1-beam \\
  \hline
  Electron position & 5\% & 6\%\\
  Mirror & 4\% & 4\% \\
  Winston cone & 2\% & 2\% \\
  Row aluminum & 3\% & 3\% \\
  Cherenkov catcher & 0\% & 1\% \\
  Others (lens, filter ...) & $<2$ \% & $<2\%$ \\ \hline
\textbf{Geometrical Acceptance} & \textbf{7.5\%} & \textbf{8.2\%} \\
  \hline
\end{tabular}
\end{center}
\caption{Geometrical acceptance systematic uncertainties for both
experimental configurations :  MF1-lab (with $^{90}Sr$ source) and
MF1-beam (CERN test beam)} \label{table-GeoAcc}
\end{table}

The others sources of systematic uncertainty come from our
reconstruction method used to extract the fluorescence
signal from the data and from the PMT calibration. We identified three
sources of errors from our reconstruction method: PMT signal fitting
procedure (DL reconstruction error), background and Cherenkov estimations.
These uncertainties vary with the setup configuration and depend on the signal intensity.

Table~\ref{table-systematics} gives our estimation of the systematics
uncertainty for dry air measurements with the MF1 setup for both laboratory
and test beam configurations.
The total absolute uncertainty is 13.2\% for MF1-lab and 13.7\%
for MF1-beam. They are dominated by PMT calibration uncertainties.

\begin{table}
\begin{center}
\begin{tabular}{|c|c|c|}
  \hline
  Errors sources & MF1-lab & MF1-beam \\
  \hline
  Geometrical Acceptance & 7.5\% & 8.2\%\\
  PMT calibration & 10\% & 10\% \\
DL reconstruction & 4\% & 3.5\% \\
CDL Simulation  & 0\% & 2\% \\
Bgd Measurement & 1\% & 2\% \\ \hline
\textbf{Systematic Error} & \textbf{13.2\%} & \textbf{13.7\%} \\
  \hline
\end{tabular}
\end{center}
\caption{Systematic uncertainties on the FLY measurement for both
experimental configurations: MF1-lab and MF1-beam.}
\label{table-systematics}
\end{table}

\section{Results}

In our air FLY model presented in section~\ref{sec-model}, the light
yield is proportional to the deposited energy in air ($E_d$).
Therefore the ratio ${FLY}\over{E_d}$ becomes independent of the
energy lost by the incoming electron. It will only vary as a
function of the atmospheric conditions (pressure, temperature and
humidity).

\begin{figure}
  \includegraphics[width=\columnwidth]{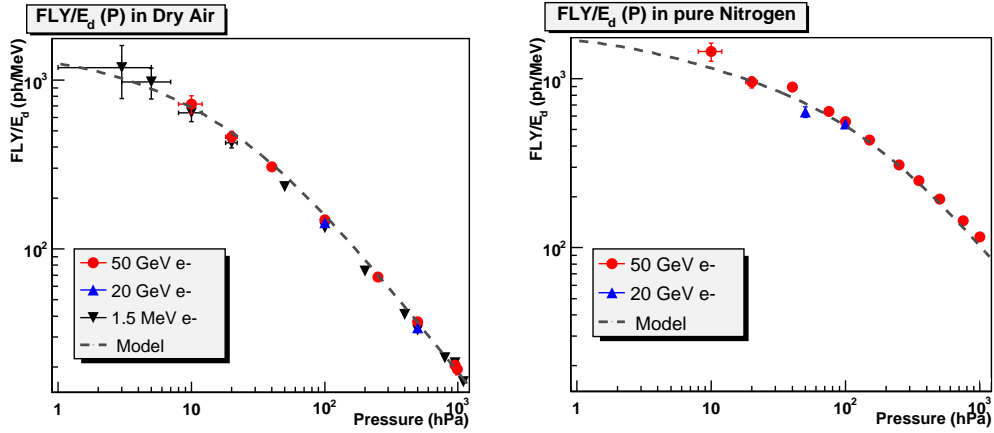}
  \caption{Ratio $FLY/E_d$ - "FLY on deposited Energy" as a function
  of the pressure, in our experimental dry air --- DA --- (left) and in
  pure nitrogen --- $N_2$ --- (right), measured for different electron
  energies: 1.5~MeV ($\blacktriangledown)$, 20~GeV ($\blacktriangle$)
  and 50~GeV($\bullet$). The dotted
  lines show the results of the fit of the data by our FLY model.}
  \label{fig-FLYonE}
\end{figure}

In the data used for this paper, the pressure is the only parameter
under consideration. Figure~\ref{fig-FLYonE} shows the
${FLY}\over{E_d}$ measured by MF1, in the wavelength range
290-440~nm, as a function of the pressure, both in our experimental
dry air (DA) and in pure nitrogen for different energies of the
incoming electron: 1.5~MeV, 20 and 50~GeV. The dotted lines
correspond to our FLY model described in section \ref{sec-model}
(equation~\ref{eq-melange}), fitted on the MACFLY data, for which
$i = N_2, O_2$. We have fitted the absolute value and the pressure
dependence of the FLY (from 3~hPa to 1100~hPa) both in our dry air (DA)
and in pure nitrogen (in same time). The fit had 3 free parameters:
the 290-440~nm integral value of pure nitrogen FLY and the mean value of
the $k_e^{N_2}$ and of the $k_e^{O_2}$.

The zero pressure extrapolation FLY obtained, respectively, for pure nitrogen
and dry air are:
\begin{equation}
    \nonumber
    {{FLY}\over{E_d}}\Big|^{N_2}_{P=0}= 1959\pm 412~ph/MeV\text{~~and~~}
    {{FLY}\over{E_d}}\Big|^{DA}_{P=0} = 1523\pm 329~ph/MeV~.
\end{equation}
The first integral value is used to compute the
$\frac{FLY}{E_d}|^{N_2}_{\lambda,P=0}$ values for the 24 $\lambda$,
according to the spectral published values~\cite{Bunner}~\cite{Nagano}~\cite{D&O}.
In the same way, the two mean $k_e^i$ values extracted from our pressure dependence
measurements are used to compute the 10 $k_e^i$ for the five exited states according
the relative values from~\cite{Pancheshnyi}~\cite{Brunet}.
Table~\ref{table-parameters} summarizes the values of all the parameters
used in our model. The $P^\prime$ are determined by the equation~\ref{eq-PDA'}
with the fitted $k_e^i$ and already published life times~\cite{Pancheshnyi}~\cite{Brunet}.

At atmospheric pressure and $T = 23^\circ$C in dry atmospheric
air, our FLY model fitted on MACFLY data gives:
\begin{equation}
    {{FLY}\over{E_d}}\Big|^{ADA} = 17.6\pm 2.3~ph/MeV~.
\end{equation}

\begin{figure}[hthb]
\begin{center}
  \includegraphics[width=10cm]{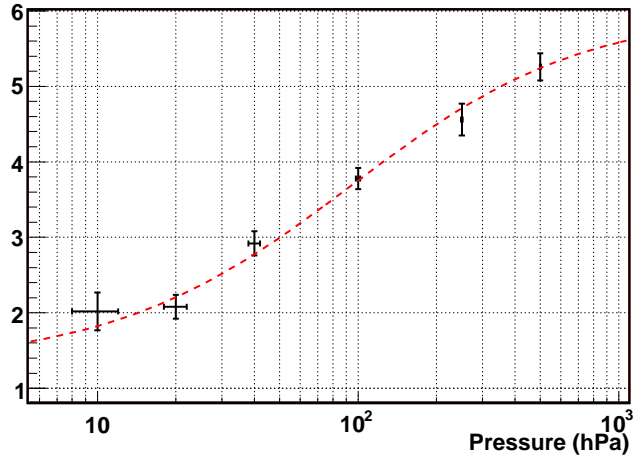}
  \caption{Ratio of the $FLY/E_d$ in $N_{2}$ and in dry air as function of pressure.
  The dotted line is the fluorescence model fit on MF1 data}
  \label{fig-N2vsAir}
\end{center}
\end{figure}

The ratio of the $FLY/E_d$ for pure nitrogen compared with the one
for dry air, is shown in figure~\ref{fig-N2vsAir} as a function of
the pressure.

At zero pressure this ratio tends to the ratio of nitrogen mass fraction
in each gas ($1/\chi_{N_2}^{DA}$ = 1.286), while at atmospheric pressure the
quenching effect of the oxygen reduces by about a factor 5.5 the FLY in pure nitrogen.

\begin{table}
\begin{center}
  \includegraphics[width=\columnwidth]{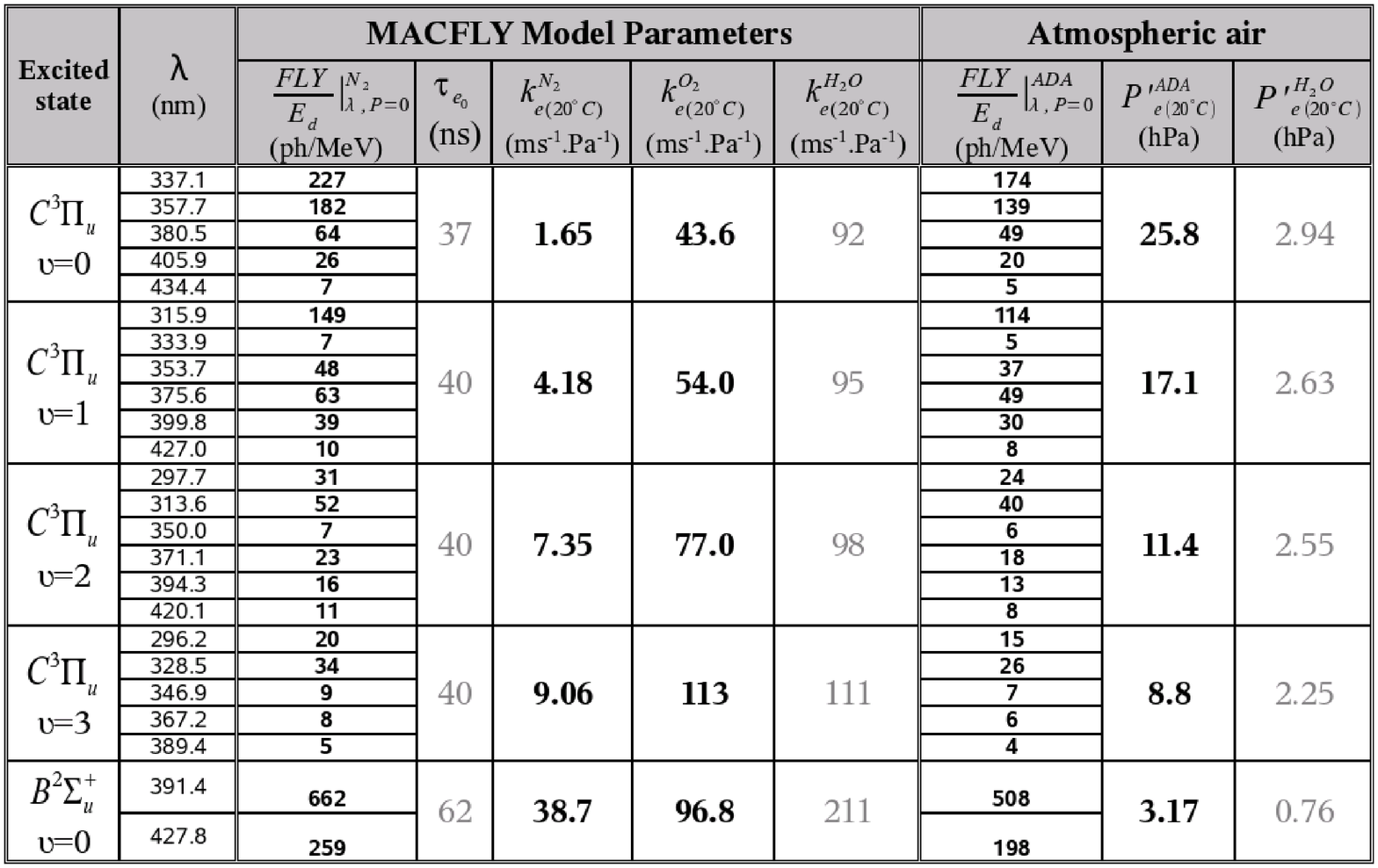}
  \caption{Parameters used in our MACFLY model describing both a
  general $N_2$,$O_2$,$H_2O$ admixture as defined in equation~\ref{eq-melange}
   and its simplification for an atmospheric
  air (equation~\ref{eq-anyairfly}). The 24 wavelengths (grouped per
  excited states) and the excited states enter the first two columns.
  The other columns contain the values of the parameters. Among them
  the bolded ones are obtained by fitting the MACFLY data.}
  \label{table-parameters}
\end{center}
\end{table}

During the data taking we measured the FLY from a "real" humid
atmospheric air (RAA). We filled MF1 chamber with the ambient
laboratory air having the following parameters: $P$=950~hPa,
$T=23^\circ$C and relative humidity 35\% (informations provided by
external probes). This corresponds to an atmospheric air (AA) with
a water vapor fraction $\mu$~=~1.05\%. We compared this
measurement with a result from our Dry Air at same pressure and
temperature. We found:
\begin{equation}\label{eq-ratiohumidair}
    \frac{FLY^{DA}}{FLY^{RAA}} = 1.11 \pm 0.07 ~,
\end{equation}
while the expected value from our model is 1.13.

\begin{table}
  \centering
    \begin{tabular}{|c|c|c|}
    \hline
    Energy & $FLY/E_d$ (ph/MeV) & $FLY/l$ (ph/m) \\
    \hline
    \hline
      1.5 MeV       & 17.0$\pm$2.3 & 3.14$\pm$0.41 \\
      20 GeV        & 17.4$\pm$2.5 & 4.22$\pm$0.61 \\
      50 GeV        & 18.2$\pm$2.5 & 4.44$\pm$0.61 \\
      {\bf All energies}  & \textbf{17.6$\pm$2.3} &  \\
     \hline
    \end{tabular}
  \caption{Values of the FLY ($FLY/E_d$ and $FLY/l$) for the
  3 energy samples, at the reference point ($P$=1013 hPa, $T=23^\circ$C,
  $\mu$=0) for atmospheric dry air (ADA). The last line shows
  the value of FLY/$E_d$ when all the data are fitted together.}
  \label{tab-fly}
\end{table}

\section{Comparison with other experiments}
To compare our results with the other experiments, we have chosen
a reference point: atmospheric dry air at $P$=1013~hPa (1~atm.)
and $T=23^\circ$C (296~K), and we express the FLY per track length
($FLY/l$) in the usual units: photon per track length ($ph/m$).
Table~\ref{tab-fly} gives values we obtained for three electron
energies: 1.5~MeV, 20~GeV and 50~GeV. These values are corrected,
according to our fluorescence model, to correspond to this
reference point.

\begin{figure}[htb]
\begin{center}
  \includegraphics[width=12cm]{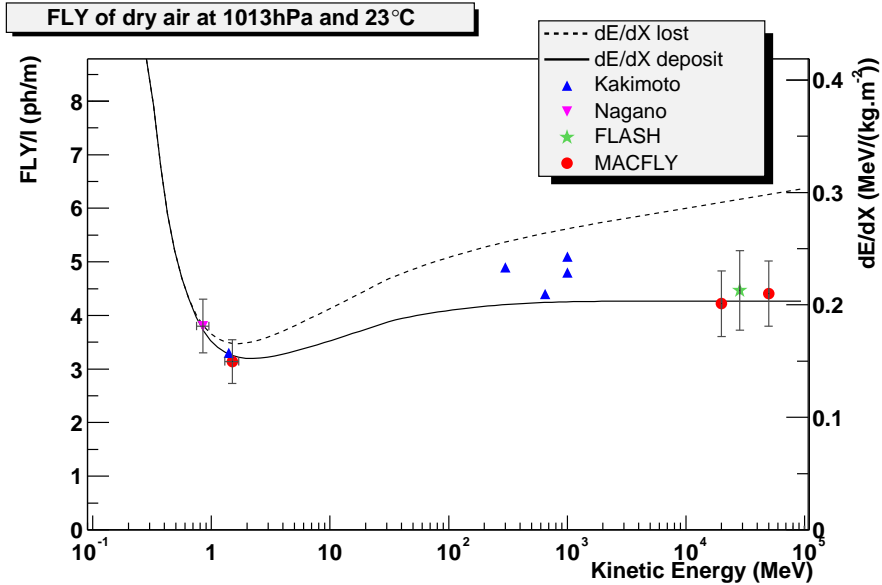}
\caption{FLY per electron track length $FLY/l$ (in photon/meter) as
a function of electron energy measured by several experiments (see
text). We also show the energy depositions (right axis): energy lost
$dE/dX$ (dashed curve) and the energy deposited $dE_d/dX$ in MF1
chamber (solid curve).}
  \label{fig-energy dep}
\end{center}
\end{figure}

In Figure~\ref{fig-energy dep} we show the $FLY/l$ measurements of
the corrected MACFLY data together with the values for others
experiments as a function of the incoming kinetic energy. On the
same figure we show the expected energy depositions in the gas. The
experimental data come from the following experiments: former
results of Nagano et al. (for 0.85~MeV electrons), the results of
Kakimoto et al., at 1.4~MeV, 300~MeV, 650~MeV and 1000~MeV, and the
recent results of the FLASH experiment~\cite{FLASH} for 28.5~GeV
electrons.

The variation of $FLY/l$ (left axis of figure \ref{fig-energy
dep}) as a function of the electron energy is compared to the
dE/dX of the electron (right axis). The beam electrons lose their
energy by ionization and by producing high energy delta and gamma
rays. The dotted line shows the total dE/dX lost by the an
electron, calculated using Berger-Seltzer formula~\cite{seltzer}
which is used by Geant4 simulation toolkit~\cite{geant4}. The
solid line indicates the energy deposited ($dE_d/dX$) in the
fiducial gas volume of the MF1 chamber,  computed with our MF1
simulation program~\cite{Colin}. The difference between the two
curves $dE/dX$ and $dE_d/dX$ reflects the energy carried away by
the high energy $\delta$-rays and $\gamma$-rays beyond the MF1
chamber. The ratio between the two scales corresponds to our FLY
model at this reference point: $FLY/E_d=17.6~ph/MeV$.

On a wide energy range (from MeV region to 50~GeV region) the MACFLY data
are well described by the deposited energy distribution. At low energy
our measurement is in agreement with results from Nagano et al. and Kakimoto
et al., following the Berger-Seltzer curve. At high energy they also agree
well with the FLASH result, showing the deposited energy behaviour of the FLY.

\section{Conclusions}

We have performed measurements of the dry air and pure nitrogen
Fluorescence Light Yields induced by single electrons of low (1.5
MeV) and high (20 and 50 GeV) energy as a function of the gas
pressure. We show that, within the experimental uncertainties, the
$FLY$ is proportional to the deposited energy in the gas,
independently of the incoming electron energy. At the reference
point: $P$=1013 hPa and $T=20^\circ$C, the $FLY/E_d$ = 17.6 $\pm$
2.3 ph/MeV.

Based on our measurements and using already published data we have
proposed a model describing, in the 390-440~nm range, the
fluorescence light yield for any air composition, as a function of
the pressure, the temperature and the water contamination.

\section{Acknowledgments}
This work has been partly supported by a fund from the Institut
National de Physique Nucl\'eaire et de Physique des Particules and
from the Joint Institute for Nuclear Research. We would like to
thank the Centre Europ\'een de Recherche Nucl\'eaire who allocated
two weeks of beam test on the SPS beam line and particularly  the
CERN PH-DT2 (Detector Technology) group. Many thanks to our
colleagues from the Alice experiment who helped and supported our
activity in the test beam area and to M. Maire for his Geant4
expertise.

\newpage
\listoffigures

\newpage
\listoftables

\end{document}